\documentclass[runningheads]{llncs}
\usepackage{hyperref}
\usepackage{tikz}
\usepackage{subcaption}
\usepackage{algpseudocode}
\usepackage{algorithm}
\usepackage{stackengine}
\usepackage{amssymb}
\usepackage{caption}

\DeclareCaptionLabelFormat{andtable}{#1~#2  \&  \tablename~\thetable}

\algrenewcommand{\alglinenumber}[1]{\tiny#1:}

\newcommand\specialcaret{%
  \stackengine{0pt}{\ \,}{\scalebox{1.1}[2]{\raisebox{-0.9ex}{\string^}}}{O}{c}{F}{T}{L}}

\begin{document}
\title{Generalized Fixed-Depth Prefix and Postfix Symbolic Regression Grammars}
%
%
\author{Edward Finkelstein\inst{1}}
\authorrunning{}
%
\institute{SDSU}
\maketitle              
\begin{abstract}
We develop faultless, fixed-depth, string-based, prefix and postfix symbolic regression grammars, capable of producing \emph{any} expression from a set of operands, unary operators and/or binary operators. Using these grammars, we outline simplified forms of 5 popular heuristic search strategies: Brute Force Search, Monte Carlo Tree Search, Particle Swarm Optimization, Genetic Programming, and Simulated Annealing. For each algorithm, we compare the relative performance of prefix vs postfix for ten ground-truth expressions implemented entirely within a common C++/Eigen framework. Our experiments show a comparatively strong correlation between the average number of nodes per layer of the ground truth expression tree and the relative performance of prefix vs postfix. The fixed-depth grammars developed herein can enhance scientific discovery by increasing the efficiency of symbolic regression, enabling faster identification of accurate mathematical models across various disciplines.

\end{abstract}

\section{Introduction}
Symbolic regression (SR) denotes finding a symbolic model $f\left(\vec{x}\right)$ that predicts a label $y$ based on an N-dimensional feature vector $\vec{x}$ while minimizing a loss metric $\mathcal{L}\left(f\left(\vec{x}\right),y\right)$. The search space contains nodes in one of the following classes:
\begin{itemize}
\item \textbf{Unary operators: } Any operator that takes 1 argument as input and outputs a value, such as $\cos$, $\sin$, $\exp$, $\ln$, $\tan$, etc.
\item \textbf{Binary operators: } Any operator that takes 2 arguments as input and outputs a value, such as $+$, $-$, $*$, $\div$, etc.
\item \textbf{Leaf Nodes: } Any of the individual features $\vec{x} = \{x_1, x_2, \ldots,x_{N}\}$ and a constant token that can be optimized with a non-linear optimization routine like L-BFGS \cite{doi:10.1137/0916069} or Levenberg-Marquadt \cite{83b09f23-b20e-3617-8f72-24765b713f7b} \cite{doi:10.1137/0111030}.
\end{itemize}
Symbolic Regression (SR) is useful for elucidating first-principles in scientific domains due to the interpretable nature of the method \cite{Angelis2023}. Nevertheless, the main challenge lies in controlling the complexity of the outputted expressions while efficiently navigating the space of possible equations. Thus, in this paper, we document prefix and postfix symbolic regression grammars that, for the first time, guarantee the generation of any possible expression of a fixed complexity, offering a robust solution for controlling expression complexity while efficiently exploring the search space. 

\subsection{Prefix and Postfix Notation in Symbolic Regression}
\textbf{Pre}fix (also known as Polish notation) expressions are written with the operators coming \emph{before} the operators, for example, \texttt{+ 1 2} represents the sum of 1 and 2. \textbf{Post}fix (also known as Reverse-Polish notation) expressions, on the other hand, have the operators coming \emph{after} the operands, for example, \texttt{1 2 +} represents the sum of 1 and 2. Both prefix and postfix remove the need for parentheses and operator precedence rules characteristic of infix notation (where operators are placed \emph{in between} operands), resulting in faster expression evaluation and less memory. An interesting question then arises: \emph{In symbolic regression, which performs better, prefix or postfix?}
\par To our knowledge, the only paper that studies the effect of prefix vs postfix expressions in SR is \cite{hemberg2008pre}.  In \cite{hemberg2008pre}, the authors compare prefix, infix, and postfix notation-based grammars on 5 bivariate equations and found that postfix performed better than prefix as the complexity increased due to the smaller percentage of invalid individuals.  
Succinctly, the 2 main limitations of \cite{hemberg2008pre} are:
\begin{enumerate}
\item The study does not consider \emph{faultless grammars}, i.e., grammars which do not produce invalids.
\item The study only considers genetic algorithms.
\end{enumerate}
In this paper, we outline faultless grammars for generating fixed-depth prefix and postfix expressions. Additionally, we benchmark the performance of 5 algorithms employing these grammars, namely, Random Search, Monte Carlo Tree Search (MCTS), Particle Swarm Optimization (PSO), Genetic Programming (GP), and Simulated Annealing (SA). 
\par Other papers \cite{lacava2021contemporary}, \cite{10.1145/3205455.3205539}, \cite{Zegklitz2021}, have performed comprehensive comparisons of different SR methods, though usually within the context of different frameworks, which could obscure otherwise very good algorithms by differences in implementation efficiency/speed. Therefore, we implement everything in one common C++ framework utilizing the Eigen template library  \cite{eigenweb}, \href{https://github.com/edfink234/Alpha-Zero-Symbolic-Regression/tree/prefix_and_postfix_cpp_implementation}{here}.

\section{Background}\label{sec:Background}
\begin{figure}[ht]

    \caption{Left: List of SR Frameworks submitted to the 2022 GECCO competition \cite{defranca2023interpretable}, their underlying expression representations, and if the expression representation was stated directly in the cited paper or implied via source-code/references. Right: Number of SR publications mentioning prefix, postfix, and acyclic graphs over time. The script to reproduce the plot is \href{https://github.com/edfink234/Alpha-Zero-Symbolic-Regression/blob/b2f7486b0797843ee363b20faa9a30677065f7b9/Figure_1/notations_pubs_counter.py}{here}.}
    \centering
    \subfloat{
    \label{GECCO_Framework_reps}
        \scalebox{0.68}{\begin{tabular*}{0.81\textwidth}{l@{\hskip 0.4in}l@{\hskip 0.4in}l} 
\hline
\textbf{Framework}&\textbf{Representation} & \textbf{Deduced}\\[0.1cm]
\hline
Bingo \cite{10.1145/3520304.3534031} &Acyclic Graph & Stated Directly\\ 
E2E Transformer \cite{kamienny2022endtoend} &prefix & Stated Directly\\ 
PS-Tree \cite{zhang2022ps} &prefix & Implied \\ 
QLattice \cite{Brolos2021AnAT} & Acyclic Graph & Stated Directly\\ 
TaylorGP \cite{10.1145/3512290.3528757} & prefix & Implied \\ 
EQL \cite{pmlr-v80-sahoo18a} & Acyclic Graph & Implied\\
GeneticEngine \cite{10.1145/3564719.3568697} & prefix & Implied \\ 
Operon \cite{10.1145/3377929.3398099} & postfix  & Stated Directly \\ 
PySR \cite{cranmer2023interpretable} & prefix & Implied \\
uDSR \cite{NEURIPS2022_dbca58f3} & prefix & Implied \\ 
$\mathrm{GP}_{\mathrm{ZGD}}$ \cite{10.1145/3377930.3390237} & prefix & Implied\\ 
NSGA-DCGP \cite{izzo2016differentiable} & Acyclic Graph & Implied \\
\hline 
\end{tabular*}}}
\subfloat
{
    \raisebox{-.5\height}{\includegraphics[width=0.43\linewidth]{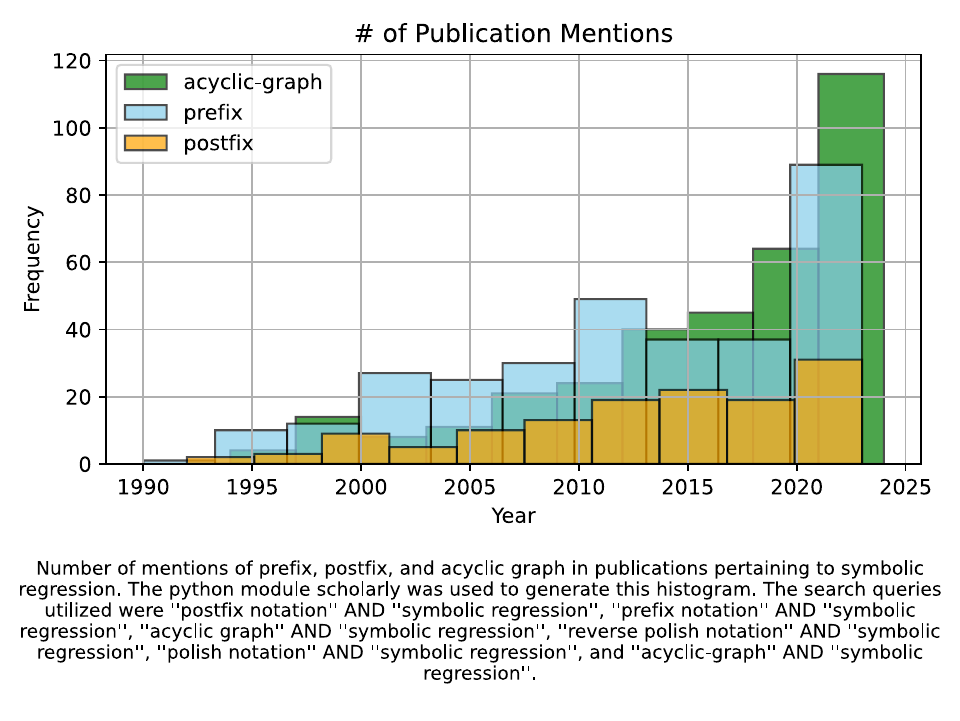}}
    \label{fig:pub_freqs_pre_post_acyc_graph}
}
\end{figure}

At the core of every SR framework lies a choice of expression representation. For example, Figure \ref {GECCO_Framework_reps} lists the frameworks considered in \cite{defranca2023interpretable} and their underlying expression representations (left) and their frequencies in SR publications over time (right).

Prefix notation and acyclic graphs are generally preferred over postfix notation for representing expressions\footnote{See Figure \ref{fig:pub_freqs_pre_post_acyc_graph} for an illustration \cite{cholewiak2021scholarly}} \cite{defranca2023interpretable}, as they allow for a natural termination condition and reduce the memory required to represent expressions with identical sub-components, respectively. The analysis in this paper is limited to prefix and postfix notations due to their similarity.


\subsection{Prefix and Postfix}
\begin{figure}[ht]
    \begin{subfigure}[b]{0.4\textwidth}
        \centering
        \begin{tikzpicture}
            \node[text width = 6 cm, align = center] at (0,0) {\includegraphics[width=\linewidth, keepaspectratio]
            {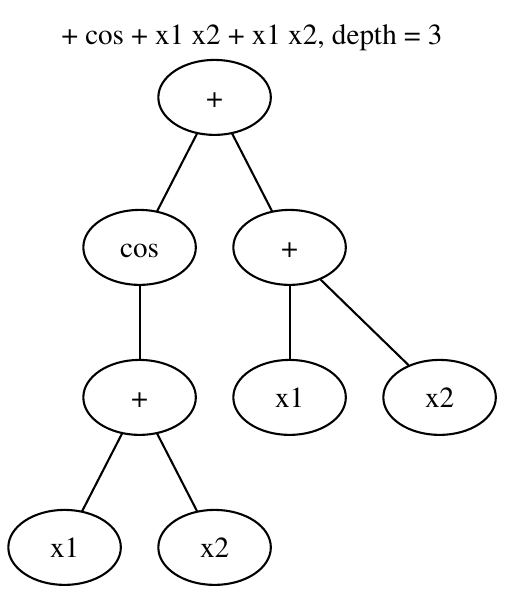}};
            \node at (-1.35, 2.4) {\textcolor{red}{\textbf{1}}};
            \node at (-2.25, 0.6) {\textcolor{red}{\textbf{2}}};
            \node at (-2.25, -1.2) {\textcolor{red}{\textbf{3}}};
            \node at (-3.15, -3) {\textcolor{red}{\textbf{4}}};
            \node at (-1.325, -3) {\textcolor{red}{\textbf{5}}};
            \node at (-0.4, 0.6) {\textcolor{red}{\textbf{6}}};
            \node at (-0.4, -1.2) {\textcolor{red}{\textbf{7}}};
            \node at (1.4, -1.2) {\textcolor{red}{\textbf{8}}};
        \end{tikzpicture} 
        \caption{prefix}
        \label{subfig:prefix_tree_example}
    \end{subfigure}
    \hspace{1cm}
    \begin{subfigure}[b]{0.4\textwidth}
        \centering
        \begin{tikzpicture}
            \node[text width = 6 cm, align = center] at (0,0) {\includegraphics[width=\linewidth, keepaspectratio]{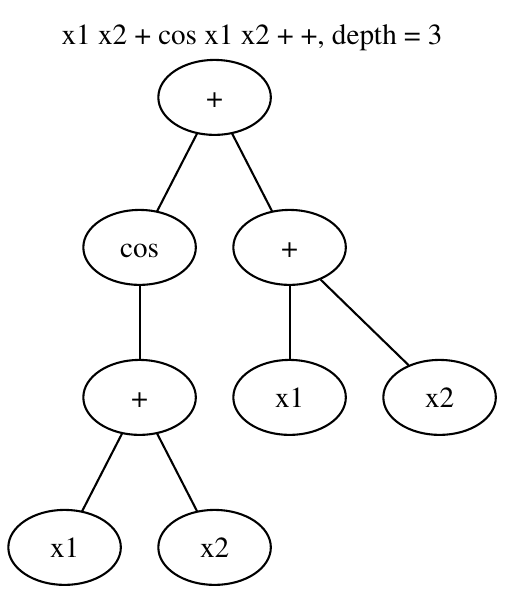}};
            \node at (-1.35, 2.4) {\textcolor{red}{\textbf{8}}};
            \node at (-2.25, 0.6) {\textcolor{red}{\textbf{4}}};
            \node at (-2.25, -1.2) {\textcolor{red}{\textbf{3}}};
            \node at (-3.15, -3) {\textcolor{red}{\textbf{1}}};
            \node at (-1.325, -3) {\textcolor{red}{\textbf{2}}};
            \node at (-0.4, 0.6) {\textcolor{red}{\textbf{7}}};
            \node at (-0.4, -1.2) {\textcolor{red}{\textbf{5}}};
            \node at (1.4, -1.2) {\textcolor{red}{\textbf{6}}};
        \end{tikzpicture}
        \caption{postfix} \label{subfig:postfix_tree_example}
    \end{subfigure}
    \caption{Prefix \& postfix representation of the infix expression $f(x_1, x_2) = \cos(x_1 + x_2) + (x_1 + x_2)$. The numbers \textcolor{red}{\textbf{1}} - \textcolor{red}{\textbf{8}} denote the order of tokens.}
    \label{fig:prefix_vs_postfix}
\end{figure}

Prefix notation entails building expressions from root nodes, while postfix notation begins expressions from leaf nodes. Figure \ref{fig:prefix_vs_postfix} illustrates the difference.

\par Prefix notation disables expansion of a completed expression tree branch without modifying a leaf node. Contrarily, postfix notation allows the tree to expand upwards from the root node, see, e.g., $\textbf{\textcolor{red}{3}} \rightarrow \textbf{\textcolor{red}{4}} $ in Figure \ref{subfig:postfix_tree_example}.

\subsection{Building Fixed-Depth Expressions}
Building fixed-depth prefix or postfix expressions requires determining the legal nodes at each step. 
\par At each step, a ``token" is added to an array. The set of selectable nodes at a given step ensures that any \emph{valid} expression can be generated with depth $N$. Tokens are appended left-to-right as they appear in prefix or postfix notation. 
\par The set of legal tokens at each step consists of unary operators and/or binary operators and/or leaf nodes of sizes $N_U$, $N_B$, and  $N_L$, respectively. $N_{T} = N_U + N_B + N_L$ denotes the total number of tokens considered. Determining which nodes are allowed in the current step depends on if the expression is prefix or postfix, explained in the remainder of this section.
\subsection{Prefix Grammar}\label{subsec:prefix_grammar}
In the first step, if the specified depth $N=0$, then the allowed tokens are the leaf nodes of size $N_L$, else, they're the operators of size $N_U + N_B$.
\par At each subsequent step, the set of allowed tokens is determined as follows:
\begin{itemize}
    \item \textbf{Unary Operators: }
    Any considered unary operator is valid if adding a unary operator to the current expression can yield an expression with depth $\leq$ the specified depth $N$. Otherwise, a unary operator is not allowed.
    \item \textbf{Binary Operators:}
    Any considered binary operator is valid if adding a binary operator to the current expression can yield an expression with depth $\leq$ the specified depth $N$. Otherwise, a binary operator is not allowed.
    \item \textbf{Leaf Nodes: }
    Any considered leaf node is valid if \emph{both} of the following conditions are false:
    \begin{enumerate}
    \item The current expression has \texttt{num\_leaves == num\_binary + 1}.
    \item Adding a leaf node results in the minimum depth being $<$ the desired depth $N$ \textbf{and} \texttt{num\_leaves == num\_binary} in the current expression.
    \end{enumerate}
\end{itemize}

\subsection{Postfix Grammar}\label{subsec:postfix_grammar}
In the first step, the only allowed tokens are the leaf nodes of size $N_L$.
\par At each subsequent step, the set of allowed tokens is determined as follows:
\begin{itemize}
\item \textbf{Unary Operators: }
Any considered unary operator is valid if $\texttt{num\_leaves} \geq 1$ \textbf{and} if adding a unary operator to the current expression can yield an expression with depth $\leq$ the specified depth $N$. Otherwise, a unary operator is not allowed. 
\item \textbf{Binary Operators: }
Any considered binary operator is valid if $\texttt{num\_binary} \neq \texttt{num\_leaves} - 1$ in the current expression. Otherwise, a binary operator is not allowed.
\item \textbf{Leaf Nodes: }
Any considered leaf node is valid if adding a leaf node results in the minimum depth of the expression being $\leq$ the desired depth $N$. Otherwise, a leaf node is not allowed.
\end{itemize}

\subsection{Conclusion}
Evidently, the postfix grammar is simpler than the prefix grammar. The main reason for this difference is that, in prefix notation, one must \emph{additionally} ensure the expression does not complete until it reaches the desired depth $N$. \par One can determine the minimum depth and completeness of a prefix/postfix token array via the stack-based approaches shown \href{https://github.com/edfink234/Alpha-Zero-Symbolic-Regression/blob/0b5b6d0b56c2d108dda023a337edeb1084436da7/PrefixPostfixSR.cpp#L393-L485}{here} and \href{https://github.com/edfink234/Alpha-Zero-Symbolic-Regression/blob/0b5b6d0b56c2d108dda023a337edeb1084436da7/PrefixPostfixSR.cpp#L487-L641}{here}, respectively\footnote{The functions \texttt{getPNDepth} and \texttt{getRPNDepth} come from \cite{77180279} and \cite{77128902}, respectively. Caching is added for optimization.}.

The next section explains the symbolic regression algorithms used in this paper in the context of the fixed-depth grammars developed in this section.

\section{Symbolic Regression Algorithms}\label{sec:SymbolicRegressionAlgorithms}

This section details the fixed-depth-grammar algorithms of this paper: Random Search, Monte Carlo Tree Search (MCTS), Particle Swarm Optimization (PSO), Genetic Programming (GP), and Simulated Annealing (SA). Given our limited computational resources, we generally opted to use hyperparameter values commonly recommended in the literature and known to perform robustly across a range of scenarios.

\subsection{Random Search}\label{subsec:RandomSearch}

Random Search, i.e., the brute-force approach, has gained more attention recently \cite{Heule2017TheSO} as it can rival GP if used with a restrictive grammar \cite{Kammerer2020} or sophisticated search strategy \cite{udrescu2020ai}.
Our random-search approach is as follows:
\par First, one starts with an empty expression list. Then, random \emph{legal} tokens are appended to the expression list until complete with depth $N$, after which the constant tokens (if any) are initialized to 1, optimized with 5 iterations of Levenberg-Marquadt, and cached as initial seeds in case the depth-$N$ expression is re-encountered. Finally, the score of the expression is computed as:

\begin{equation}
\mathrm{Score} = \frac{1}{1+ 1/N_{\mathrm{dat}}\left|\left|\hat{Y}-\vec{Y}\right|\right|^2} \qquad  (0 \leq \mathrm{Score} \leq 1), \label{eq:score_formula}
\end{equation}

where $N_{\mathrm{dat}}$ is the number of data points, $\hat{Y}$ are the predicted labels, and $\vec{Y}$ are the true labels.

\newcommand{\textunderset}[2]{\begin{tabular}[t]{@{}c@{}}#2\\[-0.3em]\scriptsize#1\end{tabular}}

\subsection{Monte Carlo Tree Search}\label{subsec:MonteCarlo TreeSearch}

Monte Carlo Tree Search (MCTS) is a popular method for navigating large game state spaces; it combines random exploration and decision-making to gather statistics and refine its search policy \cite{Silver2016} \cite{Swiechowski2023}.

In symbolic regression, MCTS has been used since \cite{CazenaveMCTS}. Recently, MCTS has seen use with policy and value estimators, e.g., in \cite{Lu2021}, Actor-Critic neural networks were used for policy and value estimation. Similarly, \cite{10.5555/3618408.3619047} leverages a neural network to output a mutation policy and an MCTS-induced value approximation. Here, we adapt the non-neural network approach of \cite{sun2023symbolic} as follows:
\par As in section \ref{subsec:RandomSearch}, an iteration begins with the set of legal tokens given the current expression list, i.e., the current state $s$. Given these tokens, we choose the best token $a$ as the first one with 0 visit counts, i.e., $N(s,a) = 0$.  If $\forall a$, $N(s,a) \neq 0$, we choose the Upper-Confidence Tree (UCT) formula action \cite{sun2023symbolic}:

\begin{equation}
\mathrm{UCT} = \textunderset{$a \in \mathcal{A}$}{$\mathrm{argmax}$}\left(Q(s,a) + c\sqrt{\frac{\ln{[N(s)]}}{N(s,a)}}\right), \label{eq:UCT_formula}
\end{equation}

where $N(s)$ is the visit count of state $s$, $c$ controls the exploration (second term in \ref{eq:UCT_formula}) exploitation (first term in \ref{eq:UCT_formula}) tradeoff and $\mathcal{A}$ is the set of all legal actions (given from the grammars in section \ref{sec:Background}) from the current state $s$. Following previous work \cite{Swiechowski2023} \cite{Auer2002}, \cite{kuleshov2014algorithms} \cite{10.1007/11871842_29}, we initialize $c \gets \sqrt{2}$. After an expression is completed and the score is computed using \ref{eq:score_formula}, the $Q$ values for all state-action pairs encountered during the generation of the expression are updated as \cite{sun2023symbolic}:

\begin{equation}
Q(s,a) = \max{\left(Q(s,a), \mathrm{score}\right)} \label{eq:update_Q_policy}
\end{equation}

where, for the first visit to $(s,a)$, $Q(s,a) \gets 0$ on the right side of equation \ref{eq:update_Q_policy}. 
\par Lastly, every $N_{\mathrm{iter}}$ iterations, if the best score has not improved, $c \gets c + \sqrt{2}$, else, $c \gets \sqrt{2}$. This method aims to balance the exploitation of a promising neighborhood of the search space while resorting to exploration when there is confidence $\propto N_{\mathrm{iter}}$ that the neighborhood has been fully exploited.

\subsection{Particle Swarm Optimization} \label{subsec:ParticleSwarmOptimization}
Particle Swarm Optimization (PSO) is a global optimization heuristic that uses ``particles'' to find the optimal value of an $\mathbb{R}^{D\in\mathbb{N}}$ objective  \cite{clerc:hal-00764996}.
\par There exists literature documenting the use of PSO in SR. The deap library implements \href{https://github.com/DEAP/deap/blob/60913c5543abf8318ddce0492e8ffcdf37974d86/examples/pso/basic.py}{general-purpose PSO} as defined in \cite{PoliOverviewPSO}, where each particle has a position representing a candidate solution and a velocity which perturbs the particle's position towards progressively better positions \cite{DEAP_JMLR2012}. In \cite{10.1007/978-3-319-70093-9_37}, candidate particles are bit-string symbolic expressions, optimized with different PSO variants. In \cite{Lu2021}, the authors use PSO to optimize the numerical constants of candidate expressions. Lastly, \cite{KARABOGA20121} uses the ``Artificial bee colony algorithm'' variant, shown to be robust on various benchmarks. Our approach is most similar to \cite{DEAP_JMLR2012} and \cite{10.1007/978-3-319-70093-9_37} and is as follows:
\par We start with 0 particles and create tokens as needed with random initial positions $U(0,1)$ and velocities $U(-1,1)$. We then round the position of the i'th particle to the nearest integer, take the modulo w.r.t. the number of allowed tokens at the current step, and select the token corresponding to the i'th particle's position. The velocity of the particle is then updated as \cite{clerc:hal-00764996} \cite{offShellPSO}:
\begin{equation}
		v \gets 0.721\cdot v + \phi_1 \cdot r_g \cdot (\mathrm{bp}_i - \mathrm{pp}_i) + \phi_2\cdot r_p \cdot (p_i - \mathrm{pp}_i) + c,
\end{equation}

where $\mathrm{bp}_i$ is the position of particle $i$ encountered in the best expression thus far, $p_i$ is the position of particle $i$ that yields the highest average score thus far, $\mathrm{pp}_i$ is the current position of particle $i$, $r_p$ and $r_g$ are random numbers $U(0,1)$, $\phi_1 \equiv 2.8$, and $\phi_2 \equiv 1.3$  \cite{offShellPSO}. The parameter $c$ is initialized to 0 and after $N_{\mathrm{iter}}$ iterations, if the best score has not improved, $c \gets U(-m, m)$, where $m \in \mathbb{N}$ starts at 1 and is increased by 1 every $N_{\mathrm{iter}}$ iterations. After an expression is completed, the best positions are updated. 

\subsection{Genetic Programming} \label{subsec:GeneticProgramming}
Genetic Programming (GP) explores the set of allowed symbolic models through operations such as crossover and mutation on an initial population \cite{manti2023discovering} \cite{PoliFieldGuideGP}, \cite{Koza1994}. Here, we employ a traditional approach similar to \cite{manti2023discovering}. First, we generate 2000 individuals using the method of section \ref{subsec:RandomSearch}. Then, we generate an additional $\approx 2000$ individuals through crossover and mutation with probabilities 0.2 and 0.8, respectively, and the best 2000 individuals of the total $\approx 4000$ individuals are passed down to the next generation. This process repeats $N_{\mathrm{gen}}$ times.
\par In this paper, the mutation and crossover operations are done directly on the expression list; \emph{no intermediary tree data structures are created}. Furthermore, the mutation and crossover operations do not alter the depth $N$ of the expression; the final depth \emph{never} changes in the algorithms detailed in this paper. 
\par The mutation procedure starts by selecting a random integer $n$ from $0$ to $N-1$, where $N$ is the fixed depth initially specified, and a random expression of depth $n$ is subsequently generated using the method described in section \ref{subsec:RandomSearch}. Then, an individual from the 2000 individuals who began the generation is randomly selected, and all of the depth $n$ sub-expressions within this randomly selected individual are identified, namely, the corresponding start and stop indices within the expression list. From all of the depth $n$ sub-expressions in the selected individual, one sub-expressions is selected at random and swapped with the randomly generated depth $n$ expression generated at the start of the mutation procedure, producing a new individual who is added to the population and whose fitness is evaluated according to \ref{eq:score_formula}.
\par The crossover procedure $\simeq$ the mutation procedure, except that, instead of generating a random depth $n$ expression, we randomly select 2 unique individuals from the initial population. We then identify all depth $n$ sub-expressions in the 2 individuals and randomly select one from each individual. Finally, we swap the 2 selected sub-expressions, producing two new individuals who are added to the population and whose scores are computed according to \ref{eq:score_formula}.

\par Both the mutation and crossover procedures require identifying all depth-$n$ sub-expressions in an expression list, in our case, without creating an intermediary tree data structure. To achieve this, we iterate over the expression list and compute the right or left grasp bound of each element, depending on whether the expression representation is prefix or postfix, respectively\footnote{\cite{3ce09117-c08b-3ddb-b2ba-3ea8005b2118} only shows the routine for the left grasp bound (top of pg. 165). Computing the right grasp bound just requires changing \texttt{*ind = *ind - 1} to \texttt{*ind = *ind + 1}.} \cite{3ce09117-c08b-3ddb-b2ba-3ea8005b2118}. Then, we compute the depth of the sub-expression between the current element's index and its grasp bound, and finally, we append these starting and stopping indices to a list if the sub-expression has depth $n$.

\subsection{Simulated Annealing} \label{subsec:SimulatedAnnealing}
Simulated Annealing is a search strategy inspired by the metallurgic annealing process for improving industrial qualities of solid alloys \cite{vanLaarhoven1987} \cite{10.1145/3449639.3459345}. 
\par Our fixed-depth approach starts by generating one random expression of depth $N$ via the method of section \ref{subsec:RandomSearch} and computing the score \ref{eq:score_formula}. We then perturb the expression in the same way as the mutation procedure in section \ref{subsec:GeneticProgramming}, except that the depth of sub-expression we swap for a random one in this section has a depth $[0,N]$ instead of just $[0,N-1]$. The perturbed expression is scored, and, if $>$ max score, we keep the perturbed expression as the current expression, else, we keep the perturbed expression with probability\footnote{Note that equation \ref{eq:prob_accept_bad_SA} can very well exceed 1; we treat this case as $P=1$.} \cite{10.1145/3449639.3459345}
\begin{equation}
		P = e^{\Delta/T}, \qquad (\Delta \equiv \mathrm{score}_{\mathrm{perturbed}} -   \mathrm{score}_{\mathrm{max}}). \label{eq:prob_accept_bad_SA}
\end{equation}
We set $T_{\mathrm{min}} = 0.012 \leq T \leq T_{\mathrm{max}} = 0.1$ and $T_{\mathrm{init}} = T_{\mathrm{max}}$ \cite{10.1145/3449639.3459345}\footnote{We set $T_{\mathrm{min}} \gets 0.012$ instead of 0.0001 as in \cite{10.1145/3449639.3459345} because of the 4-byte floating point precision of variables used in the underlying framework of this paper.}. We use the same temperature update rule as in \cite{10.1145/3449639.3459345}, namely, $T \gets rT$, where $r \equiv \left(T_{\mathrm{min}}/T_{\mathrm{max}}\right)^{1/(i+1)}$ where $i$ is the iteration number. Lastly, to escape local minima, if the best score has not improved after $N_{\mathrm{iter}}$ iterations, we update the temperature as $T \gets \min{\left(10\cdot T, T_{\mathrm{max}}\right)}$, else, we update the temperature as $T \gets \max{\left(T/10, T_{\mathrm{min}}\right)}$.

\section{Results}\label{sec:Results}
In this section\footnote{The benchmarks results in this section can be reproduced by running the script \href{https://github.com/edfink234/Alpha-Zero-Symbolic-Regression/blob/b2f7486b0797843ee363b20faa9a30677065f7b9/PrefixPostfixSR.cpp}{here} using the compilation directive given in the \href{https://github.com/edfink234/Alpha-Zero-Symbolic-Regression/blob/b2f7486b0797843ee363b20faa9a30677065f7b9/README.md}{README.md file}}, we benchmark the algorithms defined in section \ref{sec:SymbolicRegressionAlgorithms} employing the grammars defined in section \ref{sec:Background}. We set the $N_{\mathrm{iter}} \gets 50,000$ for all benchmarks\footnote{Obtained after significant pre-tuning.}.
\par First, in section \ref{subsec:HembergBenchmarks}, we perform benchmarks for the five expressions considered in \cite{hemberg2008pre}. Then, in section \ref{subsec:FeynmanBenchmarks}, we consider five of the equations in \cite{udrescu2020ai}, each with varying depth of expression tree.
\par For each configuration,
we perform 50 2 minute runs, where the MSE is sampled every 6 seconds for each run.  
This approach is similar to how the SR algorithms in \cite{defranca2023interpretable} are given a pre-specified time budget and how \cite{manti2023discovering} plots the mean and variance of the best-achieved fitnesses over time. 

\subsection{Hemberg Benchmarks} \label{subsec:HembergBenchmarks}
\begin{figure}
    \centering
    
    \begin{subfigure}{0.49\textwidth}
        \includegraphics[width=\linewidth, keepaspectratio]{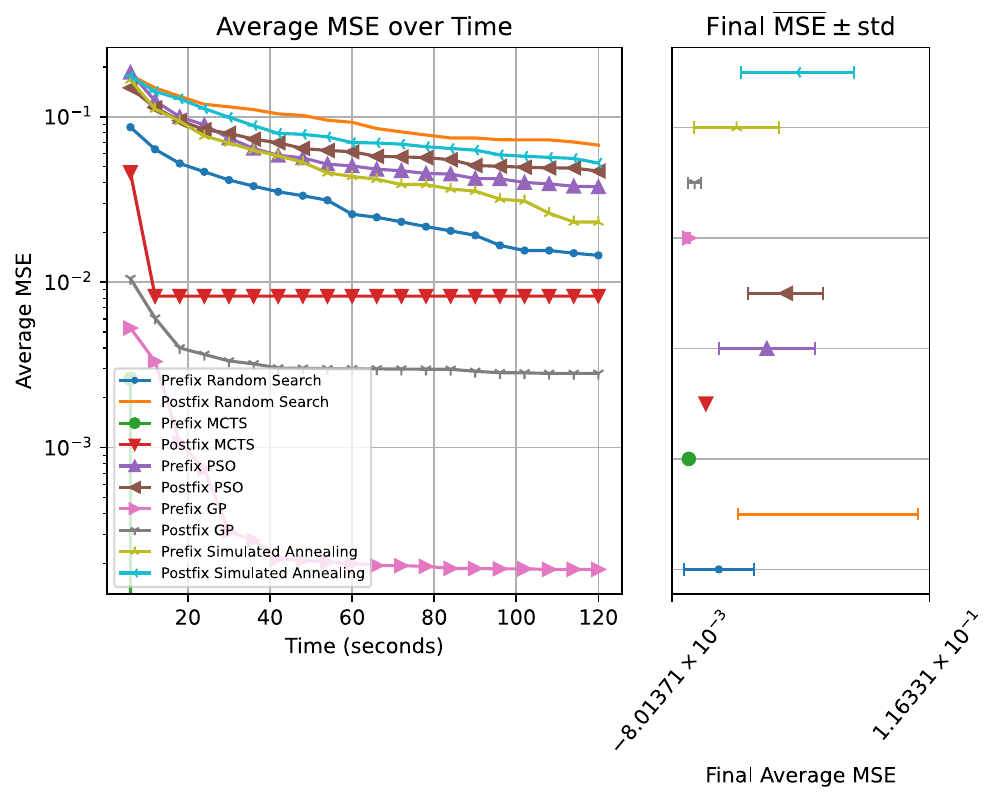}
        \caption{Hemberg 1}
        \label{subfig:hemberg_1}
    \end{subfigure}
    \begin{subfigure}[b]{0.49\textwidth}
        \includegraphics[width=\linewidth, keepaspectratio]{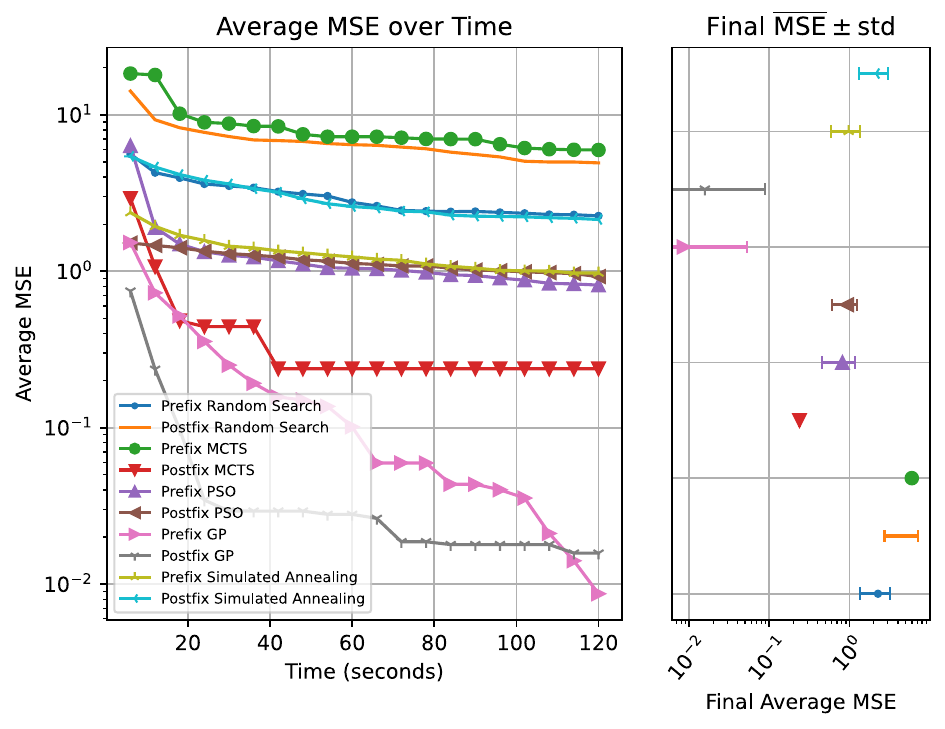}
        \caption{Hemberg 2}
        \label{subfig:hemberg_2}
    \end{subfigure}
    \vspace{0.4cm}
    
    \begin{subfigure}[b]{0.49\textwidth}
        \includegraphics[width=\linewidth, keepaspectratio]{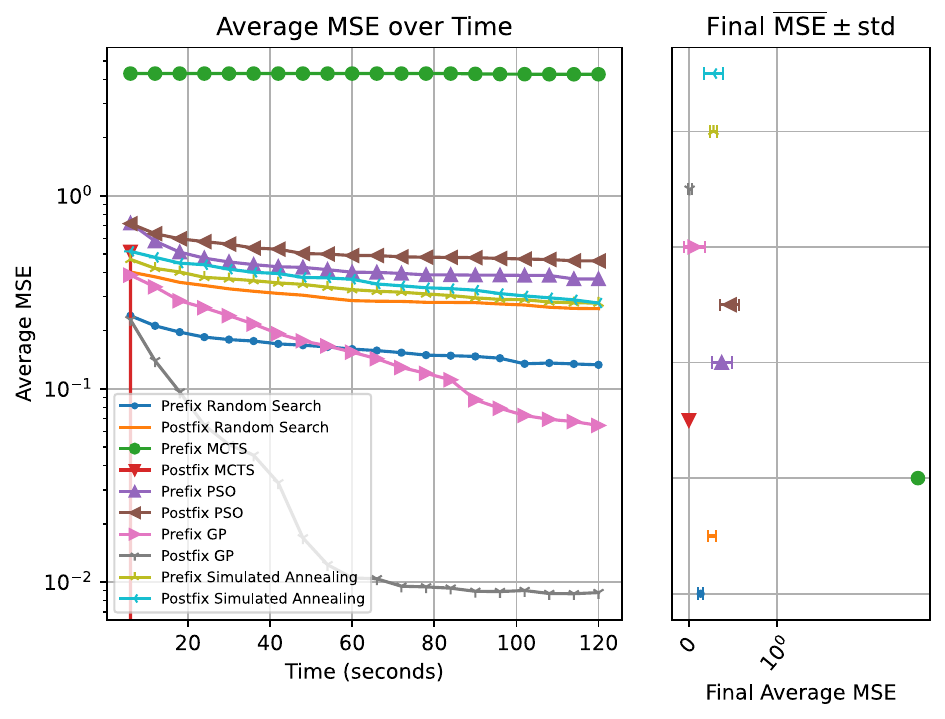}
        \caption{Hemberg 3}
        \label{subfig:hemberg_3}
    \end{subfigure}
    \begin{subfigure}[b]{0.49\textwidth}
        \includegraphics[width=\linewidth, keepaspectratio]{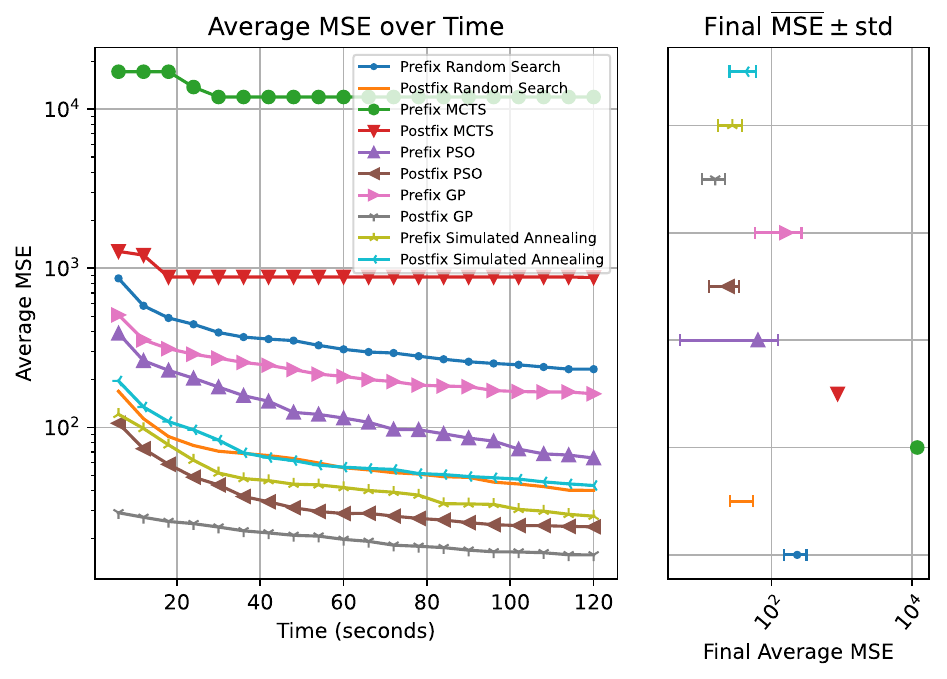}
        \caption{Hemberg 4}
        \label{subfig:hemberg_4}
    \end{subfigure}
    
    \vspace{0.4cm}
    \begin{subfigure}[b]{0.49\textwidth}
        \includegraphics[width=\linewidth, keepaspectratio]{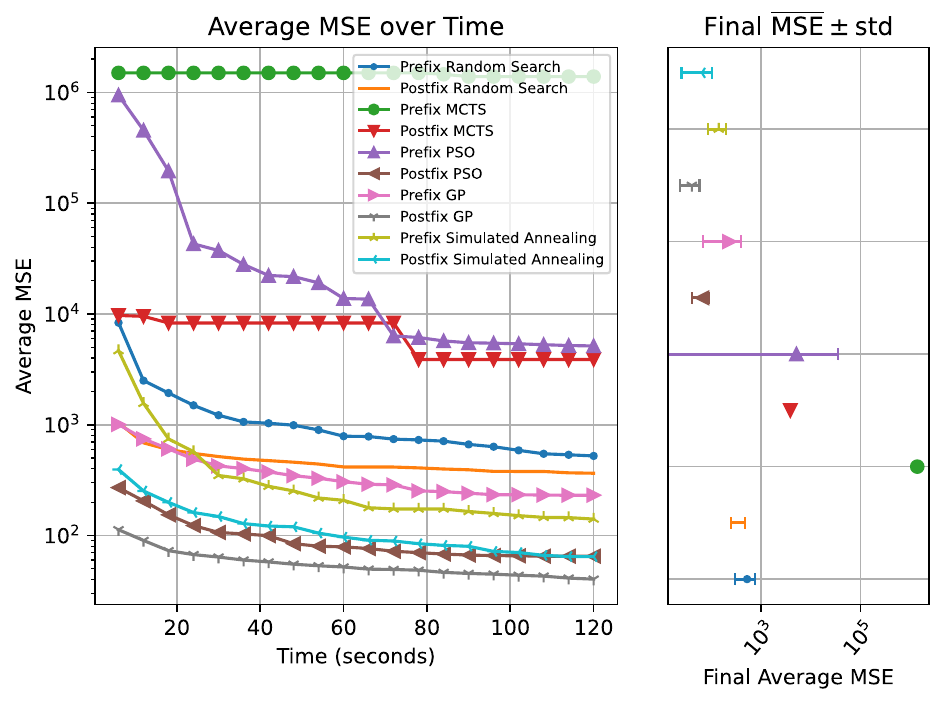}
        \caption{Hemberg 5}
        \label{subfig:hemberg_5}
    \end{subfigure}
    
    \caption{Hemberg Benchmark Equations 1-5 (from Table \ref{tab:Hemberg2008PreIP_results}). Left subplots: Average MSE over 50 runs of 2 minutes each. Right Subplots: Final Average MSE $\pm$ 1 standard deviation after 2 minutes.}
    \label{fig:Hemberg_Benchmarks}
\end{figure}

In this section, we benchmark the 5 equations of \cite{hemberg2008pre}, tabulated in Table \ref{tab:Hemberg2008PreIP_results} as well as the depth $N$ that we run the benchmarks with\footnote{In other words, we fix the tree-depth $N$ before running a particular benchmark.}. As in \cite{hemberg2008pre}, we set the ranges of $x$ and $y$ to $[-3,3]$ and sample 20 random points from the benchmark equations.  The considered operators and operands are $+$, $-$, $\cdot$, $/$, \specialcaret , and $\mathrm{const}$, $x$, $y$, respectively. The results are shown in Figure \ref{fig:Hemberg_Benchmarks}.

\begin{table} 
    \centering
    \scalebox{0.8}{
    \begin{tabular}{|l|l|l|l|l|l|}
\hline 
\# & Expression & Expression Tree & Depth $N$ & \# of Inputs & Result Plot \\ \hline 
 1 &   $8/(2+x^2+y^2)$ & \href{https://github.com/edfink234/Alpha-Zero-Symbolic-Regression/blob/13f3cc08ec72008eb735a00c14084f9b0af08293/Hemberg2008Expressions/expression_tree_Hemberg2008_expr_1.pdf}{Hemberg 1} & 4 & 2 & Figure \ref{subfig:hemberg_1} \\[0.2cm]
 2 &    $x^3\cdot(x-1) + y\cdot(y/2-1)$ & \href{https://github.com/edfink234/Alpha-Zero-Symbolic-Regression/blob/13f3cc08ec72008eb735a00c14084f9b0af08293/Hemberg2008Expressions/expression_tree_Hemberg2008_expr_2.pdf}{Hemberg 2} & 4 & 2  & Figure \ref{subfig:hemberg_2} \\[0.2cm]
 3 & $x^3/5 + y^3/2 - y - x$ & \href{https://github.com/edfink234/Alpha-Zero-Symbolic-Regression/blob/13f3cc08ec72008eb735a00c14084f9b0af08293/Hemberg2008Expressions/expression_tree_Hemberg2008_expr_3.pdf}{Hemberg 3} & 5 & 2  & Figure \ref{subfig:hemberg_3} \\[0.2cm]
  4 &   $\frac{30\cdot x^2}{(10-x)\cdot y^2}+x^4 - x^3 + \frac{y^2}{2} - y + \frac{8}{2+x^2+y^2} + x$ & \href{https://github.com/edfink234/Alpha-Zero-Symbolic-Regression/blob/13f3cc08ec72008eb735a00c14084f9b0af08293/Hemberg2008Expressions/expression_tree_Hemberg2008_expr_4.pdf}{Hemberg 4} & 9 & 2  & Figure \ref{subfig:hemberg_4} \\[0.2cm]
  5 &   $\frac{30\cdot x^2}{(10-x)\cdot y^2}+x^4 - \frac{4}{5}x^3 + \frac{y^2}{2} - 2y + \frac{8}{2+x^2+y^2} + \frac{y^3}{2} - x$ & \href{https://github.com/edfink234/Alpha-Zero-Symbolic-Regression/blob/13f3cc08ec72008eb735a00c14084f9b0af08293/Hemberg2008Expressions/expression_tree_Hemberg2008_expr_5.pdf}{Hemberg 5} & 10 & 2 & Figure \ref{subfig:hemberg_5} \\[0.2cm] \hline
\end{tabular}}
    \caption{Expressions from \cite{hemberg2008pre} considered in section \ref{subsec:HembergBenchmarks}.}
    \label{tab:Hemberg2008PreIP_results}
\end{table}

\subsection{Feynman Benchmarks} \label{subsec:FeynmanBenchmarks}
\begin{figure}
    \centering
    
    \begin{subfigure}[b]{0.49\textwidth}
        \includegraphics[width=\linewidth, keepaspectratio]{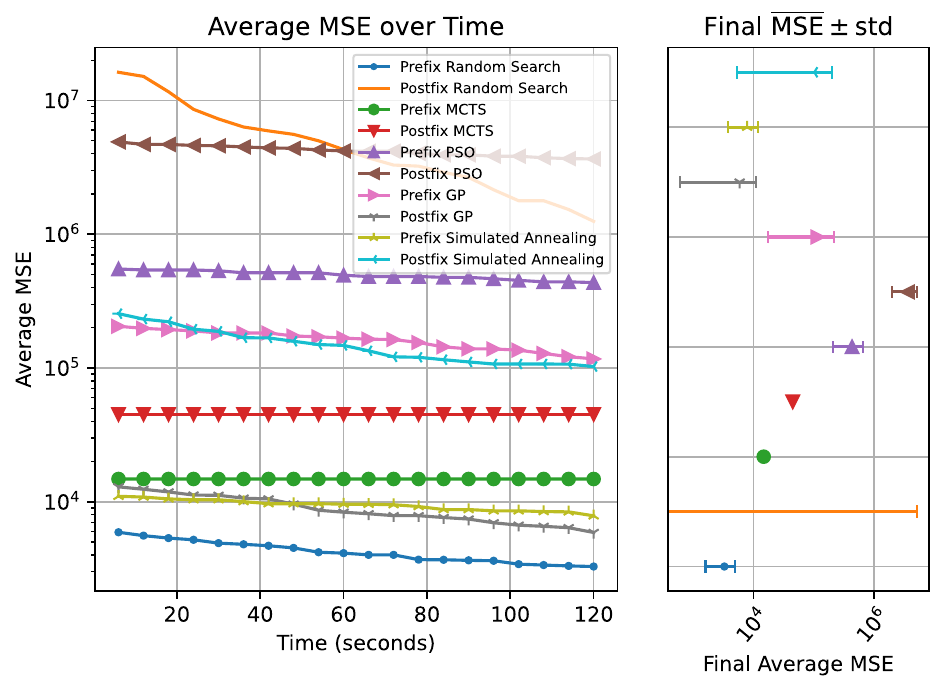}
        \caption{Feynman 1}
        \label{subfig:feynman_1}
    \end{subfigure}
    \begin{subfigure}[b]{0.49\textwidth}
        \includegraphics[width=\linewidth, keepaspectratio]{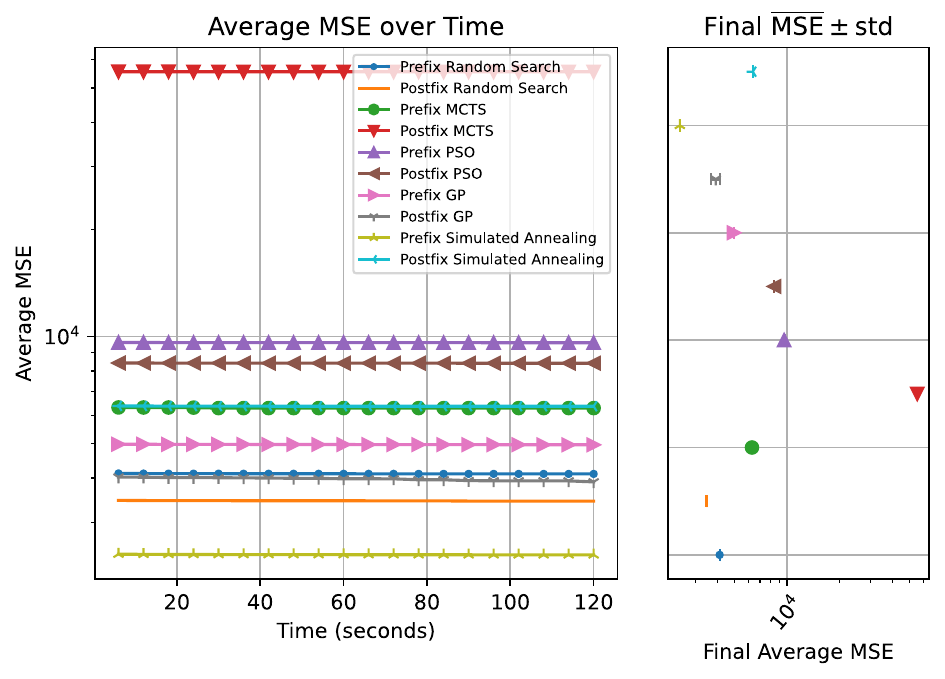}
        \caption{Feynman 2}
        \label{subfig:feynman_2}
    \end{subfigure}
    
    \vspace{0.5cm}
    
    \begin{subfigure}[b]{0.49\textwidth}
        \includegraphics[width=\linewidth, keepaspectratio]{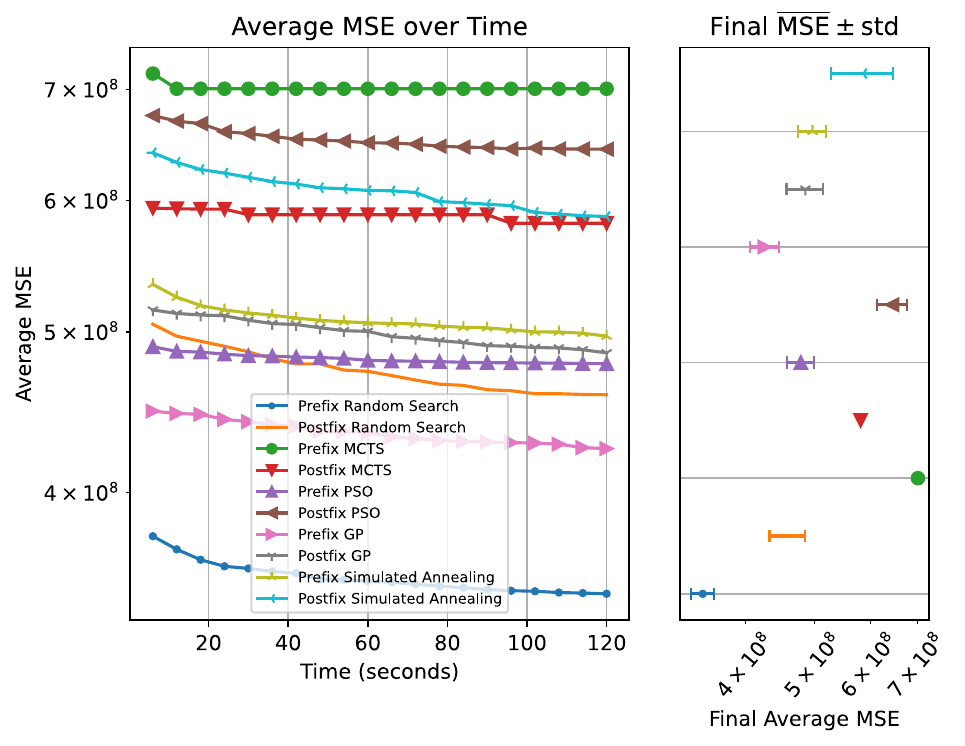}
        \caption{Feynman 3}
        \label{subfig:feynman_3}
    \end{subfigure}
    \begin{subfigure}[b]{0.49\textwidth}
        \includegraphics[width=\linewidth, keepaspectratio]{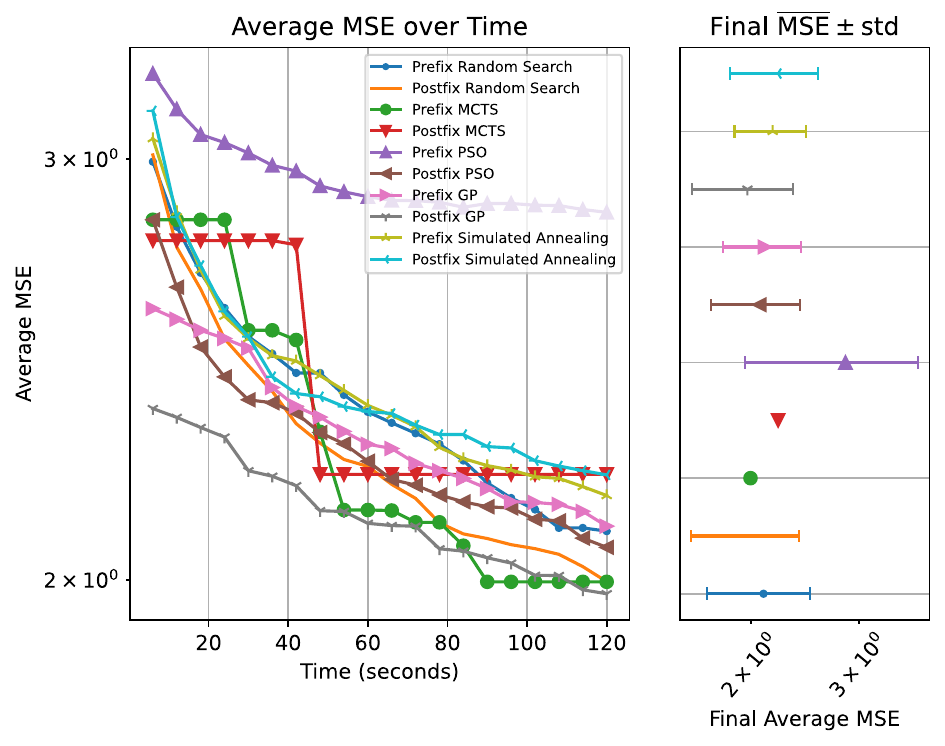}
        \caption{Feynman 4}
        \label{subfig:feynman_4}
    \end{subfigure}
    
    \vspace{0.5cm}
    
    \begin{subfigure}[b]{0.49\textwidth}
        \includegraphics[width=\linewidth, keepaspectratio]{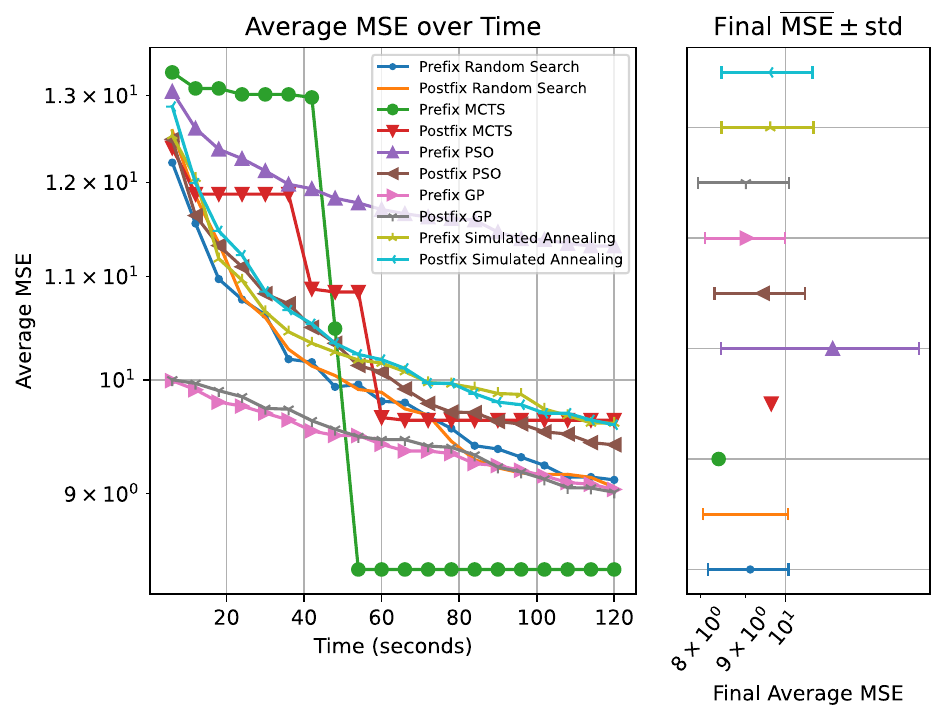}
        \caption{Feynman 5}
        \label{subfig:feynman_5}
    \end{subfigure}
    
    \caption{Feynman Benchmark Equations 1-5 (from Table \ref{tab:AI_Feynman_Benchmark_Equations}). Left subplots: Average MSE over 50 runs of 2 minutes each. Right Subplots: Final Average MSE $\pm$ 1 standard deviation after 2 minutes.}
    \label{fig:Feynman_Benchmarks}
\end{figure}

In this section, we consider 5 equations from the \href{https://space.mit.edu/home/tegmark/aifeynman.html}{Feynman Symbolic Regression Database}\footnote{For a glossary of the ``Feynman Symbolic Regression Database'' Expression Trees, see \href{https://edfink234.github.io/AIFeynmanExpressionTrees/AIFeynmanExpressionTrees/AIFeynmanExpressionTrees}{here}.}, tabulated in Table \ref{tab:AI_Feynman_Benchmark_Equations} along with the depth $N$ of their expression trees that we run the benchmarks with. As in \cite{udrescu2020ai}, we randomly sample $10^5$ data points from the benchmark expressions with input variable ranges set to $[1, 5]$. The considered operators and operands are $\sin$, $\sqrt{\phantom{1}}$, $\cos$, $+$, $-$, $\cdot$, $/$, \specialcaret, and $\mathrm{const}$, input features, respectively. The results are shown in Figure \ref{fig:Feynman_Benchmarks}.

\begin{table} 
    \centering
    \scalebox{0.82}{
    \begin{tabular}{|l|l|l|l|l|l|}
\hline 
\# & Expression & Expression Tree & Depth $N$ & \# of Inputs & Result Plot \\ \hline 
  1 &  $x = \frac{q\cdot E_f}{m\cdot \left(\omega_0^2 - \omega^2\right)}$ &  \href{https://edfink234.github.io/AIFeynmanExpressionTrees/AIFeynmanExpressionTrees/AIFeynmanExpressionTrees\#figcaption61}{Feynman 1} & 4 & 5 & Figure \ref{subfig:feynman_1}  \\[0.2cm]
   2 & $F = \frac{G\cdot m_1 \cdot m_2}{\left(x_2 - x_1\right)^2 + \left(y_2 - y_1\right)^2 +\left(z_2 - z_1\right)^2}$ & \href{https://edfink234.github.io/AIFeynmanExpressionTrees/AIFeynmanExpressionTrees/AIFeynmanExpressionTrees\#figcaption5}{Feynman 2} & 5 & 9 & Figure \ref{subfig:feynman_2} \\[0.2cm]
 3 & $A = \left(\frac{Z_1 \cdot Z_2 \cdot \alpha \cdot \hbar \cdot c}{4\cdot E_n \cdot \sin^2\left(\theta/2\right)}\right)^2$ &  \href{https://edfink234.github.io/AIFeynmanExpressionTrees/AIFeynmanExpressionTrees/AIFeynmanExpressionTrees\#figcaption101}{Feynman 3} & 6 & 7 & Figure \ref{subfig:feynman_3} \\[0.2cm]
 4 &   $f = \frac{\mu_m\cdot H}{k_b \cdot T} + \frac{\mu_m\cdot \alpha}{\epsilon \cdot c^2 \cdot k_b \cdot T} \cdot M$ & \href{https://edfink234.github.io/AIFeynmanExpressionTrees/AIFeynmanExpressionTrees/AIFeynmanExpressionTrees\#figcaption82}{Feynman 4} & 7 & 8 & Figure \ref{subfig:feynman_4} \\[0.2cm]
 5 &   $k = \frac{m \cdot k_G}{L^2}\cdot \left(1+\sqrt{\left(1+\frac{2 \cdot E_n \cdot L^2}{m\cdot k_G^2}\right)}\cdot \cos\left(\theta_1-\theta_2\right)\right)$ & \href{https://edfink234.github.io/AIFeynmanExpressionTrees/AIFeynmanExpressionTrees/AIFeynmanExpressionTrees\#figcaption102}{Feynman 5} & 8 & 6 & Figure \ref{subfig:feynman_5} \\[0.2cm] \hline
\end{tabular}}
    \caption{Expressions from \cite{udrescu2020ai} considered in section \ref{subsec:FeynmanBenchmarks}.}
    \label{tab:AI_Feynman_Benchmark_Equations}
\end{table}

\section{Remarks}

Figures \ref{fig:Hemberg_Benchmarks} and \ref{fig:Feynman_Benchmarks} show the varying performance of prefix vs postfix. To know when to use prefix and/or postfix, we build a decision tree based on the SR algorithm, the depth of the expression tree, number of input variables, and tree shape.
\par
\begin{figure}
    \centering
    \includegraphics[width=\linewidth]{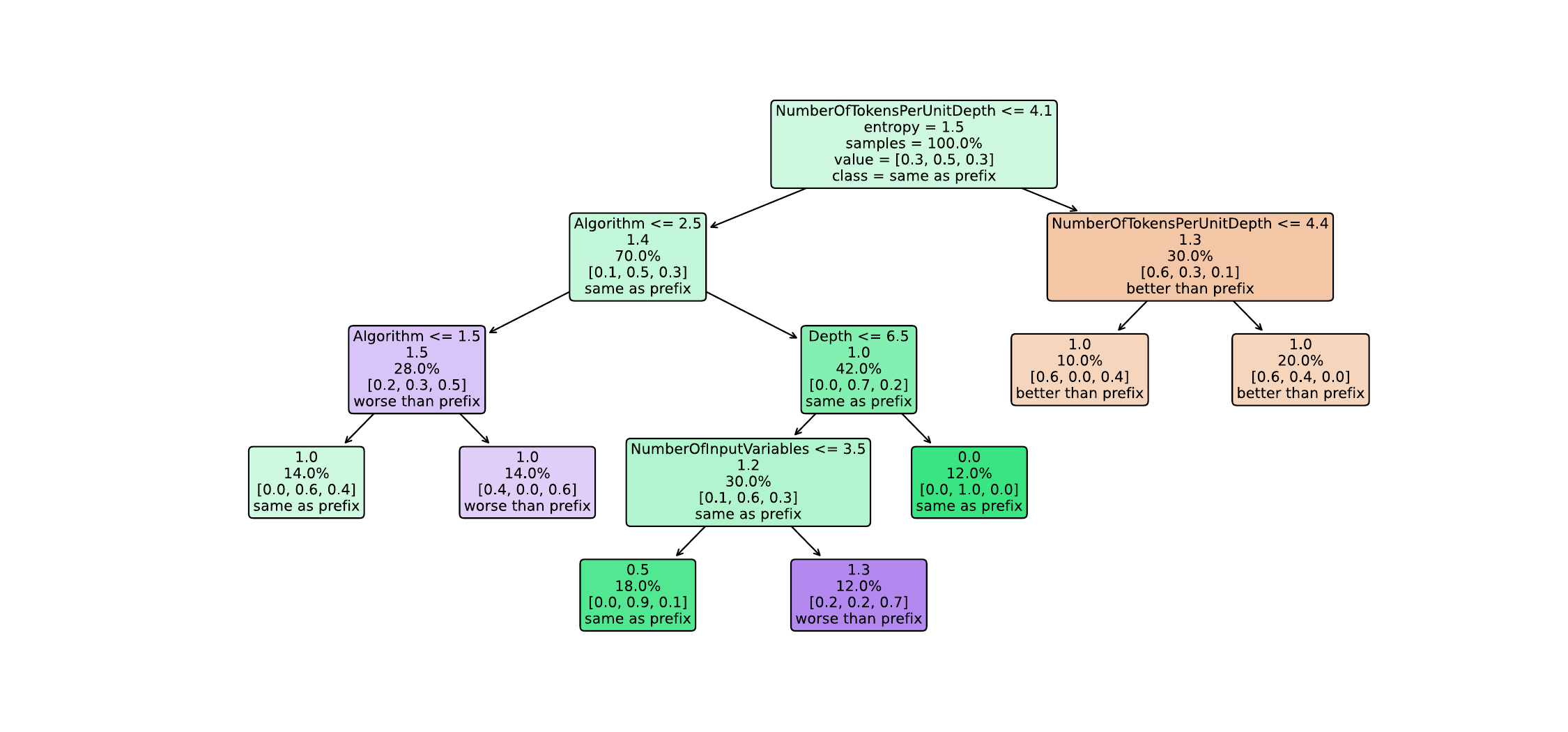}
    \caption{Decision Tree for determining how postfix will perform relative to prefix. The algorithm enumeration is \texttt{\{1: `Random Search', 2: `MCTS', 3: `PSO', 4: `GP', 5: `Simulated Annealing'\}}. This decision tree classifies the data obtained in section \ref{sec:Results} with 70 \% accuracy. The code can be found \href{https://github.com/edfink234/Alpha-Zero-Symbolic-Regression/blob/0b5b6d0b56c2d108dda023a337edeb1084436da7/PrefixPostfixDecisionTree.py}{here}. } 
    \label{fig:PrefixPostfixDecisionTree}
\end{figure}

 Figure \ref{fig:PrefixPostfixDecisionTree} suggests that the average number of nodes per layer of the ground-truth expression is the strongest indicator of the relative performance of prefix vs postfix in the symbolic regression benchmarks considered, followed by search algorithm, depth of ground-truth expression, and lastly data dimensionality.

\section{Conclusion and Outlook}
In this work, we developed string-based grammars for generating symbolic expressions. These grammars elicit faultless prefix and postfix expressions corresponding to fixed-depth trees. With these grammars, we outlined five symbolic regression algorithms and benchmarked them on expressions from \cite{hemberg2008pre} and \cite{udrescu2020ai} within a C++/Eigen framework. The results suggest the average number of nodes per layer of the ground-truth expression to be the most influential factor in determining the relative performance of prefix versus postfix in the symbolic regression benchmarks considered. Future work could expand the analysis for a larger corpus of test expressions and hyper-parameter tunings, contingent on available resources. A promising application of this work could be implementing a fixed-depth option in existing SR frameworks/grammar implementations to focus computational resources on searching the space of fixed-depth expressions instead of the larger space of depth $\leq$ max-depth expressions, enhancing efficiency and probability of scientific discovery. 

\bibliographystyle{splncs04}
\bibliography{PrefixPostfixPaper}

\end{document}